\documentclass[a4paper, 12pt]{article}
\usepackage[english]{babel}
\usepackage{graphicx}
\usepackage{amsthm}
\usepackage{amssymb}
\usepackage{amsmath,amsfonts,amstext,amssymb,amsbsy,amsopn,amsthm,eucal}
\usepackage{latexsym,amsfonts,amssymb,amsmath,amsbsy}
\usepackage{color}
\usepackage{graphics}

\newtheorem{teo}{Theorem}[section]
\newtheorem{lem}[teo]{Lemma}
\newtheorem{cor}[teo]{Corollary}
\newtheorem{rem}[teo]{Remark}
\newtheorem{rems}[teo]{Remarks}

\newenvironment{Remarks}{\begin{rems}\rm}{\end{rems}}
\newtheorem{zak'}{ '}[section]
\newtheorem{exmp}[teo]{Example}
\newenvironment{Example}{\begin{exmp}\rm}{\end{exmp}}

\newcommand{\be}{\begin{equation}}

\newcommand{\bmat}{\left( \begin{array}}
\newcommand{\emat}{\end{array} \right)}

       \title{Perturbation of eigenvalues of the Klein Gordon operators II}
\author{
  Kre\v simir Veseli\'c\thanks{Fernuniversit\" at Hagen, Fakult\" at
  f\" ur Mathematik und Informatik, Postfach 940, D-58084 Hagen, Germany, 
    e-mail: Kresimir.Veselic@FernUni-Hagen.de.}}
\date{ }
\begin{document}
\maketitle
\begin{abstract} 
We give estimates for the changes of the eigenvalues of the
Klein Gordon operator under the change of the potential.
In some relevant situations we improve the existing estimates.
We test our results on some exactly solvable models (Coulomb potential,
Klein-Gordon oscillator).
\end{abstract}

\section{Introduction and preliminaries}
\setcounter{equation}{0}
The abstract time independent Klein-Gordon equation reads formally
\begin{equation}\label{KG}
(U^2 - (\lambda - V)^2)\psi = 0 
\end{equation}
where \(U\), (\(U^2\) usually meaning the kinetic  plus mass energy) is 
selfadjoint and positive definite operator in a Hilbert
space \(\mathcal{X}\) and \(V\) (the potential) 
symmetric and in some sense dominated by 
\(U\) and \(\lambda\) is the eigenvalue parameter.
The most interesting application is the standard Klein-Gordon equation with
\(\mathcal{X} = L_2(\mathbb{R}^3)\) and
\begin{equation}\label{standardU2}
U^2 = c^2p^2 + mc^2,\quad p = (-ih\nabla - \frac{e}{c}A(x))^2,\quad
V = V(x),\quad x \in \mathbb{R}^3
\end{equation}
where \(h,c\) are the common physical constants whereas the mass term \(m\)
and the magnetic potential \(A\) may be position-dependent. Or else, 
\(U^2\) may be some other elliptic differential operator.

Typically there is a spectral gap around zero, then isolated eigenvalues 
of finite multiplicity appear which may either be bounded by the spectral continuum
reaching to \(\pm \infty\) or else the whole spectrum is discrete again approaching
\(\pm \infty\). Of some interest could be also the case where
\(U,V\) are finite matrices.
The most interesting eigenvalues are those around zero.\\

The aim of this paper is to prove sharp perturbation estimates for
discrete eigenvalues \(\lambda\) under the change of \(V\) and \(U\).
The main technical tool is the monotonic dependence of the eigenvalues 
as functions of the potential. In fact, once one has monotonicity, then
for the perturbed potential \(\tilde{V} = V + \delta V\) 
(with \( \delta V\) bounded)
we would have
\begin{equation}\label{VVtilde}
V - \inf\sigma(\delta V) \leq \tilde{V} \leq V -\sup\sigma(\delta V)
\end{equation}
which immediately implies the perturbation bound
\begin{equation}\label{ll'+-}
\lambda - \inf\sigma(\delta V) \leq \tilde{\lambda} \leq 
\lambda + \sup\sigma(\delta V) 
\end{equation}
where \(\lambda,\tilde{\lambda}\) is the corresponding eigenvalue, respectively. 
In particular,
\[
\lambda - \|\delta V\| \leq \tilde{\lambda} \leq 
\lambda + \|\delta V\|
\]
 While such monotonicities are plausible for the Schroedinger and 
Dirac operators, the Klein-Gordon case is less simple, because
the potential \(V\) enters (\ref{KG}) as a quadratic polynomial.
To explain this we give a simple heuristic derivation. 
Let \(V = V_t\) be real analytic and produce \(t\)-dependent eigenvalue
\(\lambda_t\) and eigenvector \(\psi_t\) then by differentiating
(\ref{KG}) we obtain 
\begin{equation}\label{nu'1}
\lambda_t' = \frac{(V_t^2\,{'}\psi_t,\psi_t) - 
2\lambda_t(V_t'\psi_t,\psi_t)}{2((V_t - \lambda_t)\psi_t,\psi_t)}
\end{equation}
(here \('\) stays for the derivative). Supposing \(\lambda_t\) to be, say,
positive (\ref{nu'1}) will yield monotonicity, if 
\(V_t\) is negative semidefinite --- which is a notable restriction. Note
also that growing \(V_t\) will produce falling \(V_t^2\) only if 
\(V_t\) all commute (which is secured with multiplicative potentials, but
not generally).

A derivation to this effect was
produced in  \cite{Hall3}.\footnote{This kind of monotonicity seems to be
observed first in \cite{Najman1983}. The result in \cite{Najman1983}
was stated under similar restrictive conditions, but it seems to us that
the proof offered there has some gaps we were unable to fill.} Here 
we will make this result rigorous
and more precise. We will also enlarge its validity and then apply it
to obtain eigenvalue perturbation bounds.
More specifically we will
\begin{enumerate}
\item[(i)] prove that analytic dependence of \(V\) on a parameter implies
the same for the eigenvalues and eigenvectors, not just locally, but such as 
to produce eigenvalue bounds. In fact, analysing quadratic 
eigenvalue equation (\ref{KG}) involves unbounded non-selfadjoint phase 
space Hamiltonians which are symmetric with respect 
to an indefinite scalar product so some extra care has to be taken,
\item[(ii)] make sure that the desired monotonicity, as well as 
the bounds of the type (\ref{ll'+-}) hold for increasingly 
ordered eigenvalues with their multiplicities, just as is obtained
by standard minimax arguments, (this is not quite trivial
even with common selfadjoint operators if one considers the discrete
eigenvalues in gaps of the essential spectrum, see \cite{Veselic-selfa})
\item[(iii)] weaken the condition of positivity 
of the considered eigenvalues
into the more natural, so-called {\em plus} property 
(which will, roughly speaking,
cover all eigenvalues coming from the upper continuum as long as they do
not clash with those coming from below). Unfortunately we were
able to do this only partially, as yet, for instance for
potentials of the form \(tV\). 
\item[(iv)]  also weaken the non-positivity
of the potential \(V\) because even for non-positive potentials
the two-sided inclusion (\ref{VVtilde}) requires monotonicity for
slightly
indefinite potentials.
\item[(v)] give a collection of mostly exactly solvable examples
illustrating the estimates. A particularly interesting case
will be that of perturbed Coulomb potential, where the local deformation
\(\delta V\) with
\[
|\delta V(x)| \leq \beta\frac{1}{|x|},\quad \beta < 1
\]
leads to particularly tiny bound for the perturbed eigenvalue
\(\lambda'\)
\begin{equation}\label{tiny}
\lambda(1 + \beta) \leq \lambda' \leq \lambda(1 + \beta)
\end{equation}
where \(\lambda(t)\) is the corresponding eigenvalue corresponding 
to the potential \(-t/|x|\) and is given by an explicit formula.
We will also make comparison with the bounds obtained recently in \cite{NakicVeselic2020}
and show that our bounds complement the ones from \cite{NakicVeselic2020}.
\end{enumerate}

Another example will be that of the Klein-Gordon oscillator in
\(L_2(\mathbb{R}^n)\) with \(U^2 = - \Delta + x^2\) and \(V = 0\)
and then the bounds (\ref{ll'+-}) will actually hold for both sides
of the spectrum.

The plan of the paper is as follows. In Section 1. we give definitions
and fundamental properties of the Hamiltonians considered, in Section 2.
we derive the mononotonicity and in Section 3. give the resulting 
sharp eigenvalue bounds
of the type (\ref{ll'+-}), (\ref{ll'}) together with illustrating examples.
\section{The Hamiltonian formulation and analyticity}
The eigenvalue analysis necessitates rewriting a quadratic eigenvalue problem
as a linear one with 'doubled dimension'. By setting
\[
\psi_1 = \psi,\quad \psi_1 = (\lambda I - V)\psi
\]
we arrive at the eigenvalue equation \(K\psi = \mu\psi\)
with
\begin{equation}\label{K_formal}
K =
\left[\begin{array}{cc}
V      &  I \\
U^2    &   V\\
\end{array}\right].
\end{equation}
that is,
\(K = JL\) where \(L\)
\begin{equation}\label{L_formal}
L = JK =
\left[\begin{array}{cc}
U^2    &  V\\
V     &  I \\
\end{array}\right]
\end{equation}
is again formally Hermitian. So, \(K\) is \(J\)-Hermitian.
Our general assumption is 
\begin{equation}\label{condition_on_V}
\|(V - \mu I)U^{-1}\| < 1,
\end{equation}
for some real \(\mu\). This commonly used condition insures reasonable 
spectral properties of the operator \(K\), see e.g. \cite{NakicVeselic2020} and the 
literature cited there.\footnote{The Hamiltonian considered there is not the
one from the present paper but the eigenvalues and their multiplicities
are the same.} The set \(\mathcal{I}\) of all such 
\(\mu\) is obviously
an open interval and we shall call it {\em the definiteness interval}.
Under this condition \(L\) is rigorously defined by means of quadratic forms
in the factorisation
\begin{equation}\label{L_factorised}
L - \mu J =
\left[\begin{array}{cc}
U    &  0\\
0     &  I \\
\end{array}\right]
\left[\begin{array}{cc}
I    &  \overline{U^{-1}(V - \mu I)}\\
(V - \mu I)U^{-1}     &  I \\
\end{array}\right]
\left[\begin{array}{cc}
U    &  0\\
0     &  I \\
\end{array}\right]
\end{equation}
which is selfadjoint positive definite (being a symmetric product of three
such factors). Thus we obtain a \(J\)-selfadjoint operator 
\begin{equation}\label{K}
K = JL,\quad
J =
\left[\begin{array}{cc}
0    &   I\\
I     &  0 \\
\end{array}\right]
\end{equation}
 
that is, a selfadjoint operator with respect to the indefinite scalar product
\[
[\psi,\phi] = (J\psi,\phi).
\]
These operators have
 real spectrum and a rich spectral calculus,
see \cite{Langer82}. By our condition (\ref{condition_on_V}) 
we have
\[
\sigma(K) = \sigma_-(K) \cup \sigma_+(K),\quad
\sigma_-(K) < \mathcal{I} < \sigma_+(K).
\]
Moreover, as it is readily seen the eigenvalues on the right/left from
\(\mu\) have \([\cdot,\cdot]\)-positive/negative eigenvactors, and will 
be called {\em plus/minus-eigenvalues}, respectively. Also obvious is the fact
that all eigenvalues are semisimple.

The spectra of the operators \(H,K\) are connected with those of the
quadratic families like (\ref{KG}). 

The following Facts were shown in \cite{Veselic1983}, 
Thms 2.4 and 4.2.
\begin{enumerate}
\item The sesquilinear form 
\begin{equation}\label{q_sesquilinear}
q_\mu(\psi,\phi) = (U\psi,U\phi) - ((V - \mu I)\psi,(V - \mu I)\phi) 
\end{equation}
is closed and sectorial for every \(\mu \in \mathbb{C}\) as defined on
\(\mathcal{D}(U)\) and it generates a closed sectorial operator
\(Q_\mu\) whose domain \(\mathcal{D}(Q_\mu)\) is independent of 
\(\mu\).
\item
\begin{equation}\label{Q_mu_lambda}
Q_\mu = Q_\lambda + 2(\mu - \lambda)V + (\mu^2 - \lambda^2)I
\end{equation}
for any \(\mu,\lambda\).
\item Denoting by \(\rho\) the set of \(\mu\)-s for which \(Q_\mu^{-1}\)
is everywhere defined and bounded and by \(\sigma\) its complement
we have
\[
\rho = \rho(K),\quad \sigma = \sigma(K).
\]
\item \label{fact_eigen}
\[
\psi \in \mathcal{D}(Q_\mu),\quad Q_\mu \psi = 0
\]
is equivalent to
\[
\left[\begin{array}{cc}
\psi  \\
\phi   \\
\end{array}\right]
\in \mathcal{D}(H)
\]
for some \(\phi\) and
\[
H
\left[\begin{array}{cc}
\psi  \\
\phi   \\
\end{array}\right]
=
\mu
\left[\begin{array}{cc}
\psi  \\
\phi   \\
\end{array}\right].
\]

\end{enumerate}

\section{Monotonicity}
We will show that the eigenvalues depend monotonically on the
potential.

In order to compare the eigenvalues by means of analytic perturbations
we will linearly connect them as
\begin{equation}\label{connect}
 V_t = tV_1 + (1 - t)V_0 =
V_0 + t\delta V,\quad \delta V = V_1 - V_0  \geq 0
\end{equation}
\[
0 \leq t \leq 1.
\]
\begin{lem} \label{lemma} Let \(V_0,V_1\) be symmetric and 
\(\|(V_0 - \mu I)U^{-1}\|, \|(V_1 - \mu I)U^{-1}\| \leq \beta < 1\)
for some real \(\mu\).
Then 
\[
\|(V_t - \mu I)U^{-1}\| \leq \beta.
\]
\end{lem}
{\bf Proof.} It is sufficient to consider \(\mu = 0\).
We have
\[
(\|V_t\psi\|^2 = t^2(V_1\psi,V_1\psi) + t(1 - t)(V_1\psi,V_0\psi) +
t(1 - t)(V_0\psi,V_1\psi)
+ (1 - t)^2(V_0\psi,V_0\psi) 
\]
\[
\leq
t^2\|V_1\psi\|^2 + 2t(1 - t)\|V_1\psi\|\|V_0\psi\| + (1 - t)^2\|V_0\psi\|^2
\]
\[
\leq
t^2\beta\|U\psi\|^2 + 2\beta t(1 - t)\|U\psi\|\|U\psi\| + (1 - t)^2\beta
\|U\psi\|^2
= \beta(t^2 + 2t(1 - t) + (1 - t)^2)\|U\psi\|^2 
\]
\[
=  \beta\|U\psi\|^2.
\]
Q.E.D.\\

By the preceding lemma the operator \(K_t\) 
constructed with \(V_t\) via (\ref{L_factorised}), (\ref{K}) 
has the same properties as  \(K\).
In particular
\begin{equation}\label{gap}
-\frac{\|U^{-1}\|^{-1}}{1 - \beta} \leq \sigma(K_t) - \mu 
\leq \frac{\|U^{-1}\|^{-1}}{1 - \beta}.
\end{equation}
\begin{teo}\label{monotone1}
Let \(V_0,V_1,\delta V\) be as in (\ref{connect}) 
and let \(V_0,V_1\) commute, satisfy
\(\|(V_0 - \mu I)U^{-1}\|, \|(V_1 - \mu I)U^{-1}\| \leq \beta < 1\)
 and be ordered as
\begin{equation}\label{order}
 (V_1\psi,\psi) \geq (V_0\psi,\psi) \geq -M(\psi,\psi)
\end{equation}
where \(M\) is the right hand side of (\ref{gap}).

Suppose also that \(\delta V^2\) and \(\delta VV_0\) are defined on
\(\mathcal{D}(U)\). 
Denote by \(K_0,K_1\) the corresponding operators from (\ref{L_factorised}),
(\ref{K}).
Assume that the top of the minus-spectrum of \(K_0\) consists of 
discrete eigenvalues, counted with their (finite) multiplicities
\[
\lambda_1^{(0)} \geq \cdots  \geq \lambda_n^{(0)} > \tilde{\lambda}
\]
such that the negative part of \(\sigma_{ess}(K_t)\) stays left from
\(\tilde{\lambda}\) for \(t \in [0,1]\), where the operator \(K_t\)
belongs to the potential \(V_t\) from (\ref{connect}).
 Then the top of 
the minus-spectrum of \(K_1\) also consists of 
discrete eigenvalues
\[
\lambda_1^{(1)} \cdots \geq \lambda_n^{(1)} \geq ...
\]
and we have
\[
\lambda_i^{(1)} \geq \lambda_i^{(0)}.
\]
\end{teo}

Note that the last assumption in the preceding theorem is automatically 
fulfilled if the essential spectrum does not move at all with 
\(V + t\delta V\), for instance, if both \(V\) and \(\delta V\) are
either \(U\)-compact or \(U^2\)-compact  (see \cite{Najman1983} or
\cite{LangerTretter2006}).

{\bf Proof.} By the preceding lemma the operator \(K_t\)
constructed with \(V_t\) via (\ref{K_formal}) has the same properties as  \(K\)
and it is analytic in \(t\) on a region containing the closed interval
\([0,1]\). This is best seen in computing the inverse of \(L_\mu\) factorised
as in (\ref{L_factorised}), replacing \(V\) by \(V_t\) and 
using the obvious Neumann expansion
for the inverse of the middle term in (\ref{L_factorised}).
Now, as it was shown in \cite{Vespseudo} the analyticity properties
of the isolated eigenvalues are similar as with standard selfadjoint
analytic families. More precisely, let \(\lambda \geq \tilde{\lambda}\) 
be any discrete
minus-eigenvalue of \(K_t\) for some \(0 \leq t_0  \leq 1\). Then the spectrum
of \(K_t\) in a neighbourhood of \(t_0\) is represented by one or several
--- according to the multiplicity of \(\lambda(t_0)\) --- analytic functions
which, together with their J-orthonormal eigenvectors, 
can be analytically continued as long as
\begin{itemize}
\item they are separated from the continuous spectrum and
\item they do not meet a plus-eigenvalue.
\end{itemize}
(The analyticity is not destroyed even at the places where the 
eigenvalues cross.)
Both conditions are fullfilled by our assumptions. Indeed, the 
second condition is insured by the global estimate
\(\|V_tU^{-1}\| < 1\). This estimate together with the non-intruding
insures the first condition by the assumed condition 
\[
\lambda_n^{(1)} \geq \tilde{\lambda} 
\]
for some \(\lambda\). 
Consider any such analytic eigenpair \(\lambda(t),\Psi(t)\):
\begin{equation}\label{analyticEVP}
K_t\Psi(t) = \lambda(t)\Psi(t).
\end{equation}
This equation is equivalent to
\begin{equation}\label{QEP}
Q_{\lambda,t}\psi(t) = 0
\end{equation}
where
\[
\psi(t) =  \psi_1(t),\quad
\Psi(t) = 
\left[\begin{array}{c}
\psi_1(t)\\
\psi_2(t)\\
\end{array}\right]
\]
and \(q_{\lambda,t}\), \(Q_{\lambda,t}\) are defined by 
(\ref{q_sesquilinear}) with
\(V\) replaced by \(V_t\) from (\ref{connect})
--- see Fact \ref{fact_eigen} above.
Looking for monotonicity  we will  have to differentiate the
eigenvalue equation (\ref{QEP})
with respect to \(t\). This is certainly possible
because by Fact \ref{fact_eigen}  \(\psi = \psi(t)\) appearing there is 
just the first component of \(\Psi(t)\).
In order to do this it will be convenient to know that the domain
\(\mathcal{D}_t\) does not depend on \(t\) either and to have an
explicit operator
expression for \(Q_{\lambda,t}\).

Now replace in (\ref{q_sesquilinear}), (\ref{Q_mu_lambda}) \(V\) by
\(V_t = V_0 + t \delta V\) thus obtaining
\begin{equation}\label{q_mu_t}
q_{\lambda,t}(\psi, \phi) = (U\psi, U\phi) - ((\lambda - V_0 - t\delta V)\psi, 
(\lambda - V_0 - t\delta V)\phi)\quad \psi, \phi \in \mathcal{D}(U).
\end{equation}
and analogously for \(Q_{\lambda,t}\). 
Let now \(\psi\in \mathcal{D}(Q_{0,0})\) then 
\[
(U\psi, U\phi) - (V_0\psi,V_0\phi) = (Q_{0,0}\psi,\phi)
\]
and
\[
q_{\lambda,t}(\psi, \phi) = (Q_{0,0}\psi,\phi) -
\]
\[
- \lambda^2(\psi,\phi) 
- t^2(\delta V\psi,\delta V\phi) + 2\lambda(V_0\psi,\phi) 
+ 2\lambda t(\delta V\psi,\phi) - t(\delta V\psi,V_0\phi) 
- t(V_0\psi,\delta V\phi).
\] 
Using the fact that \(V_0,\ \delta V\) are commuting and that
the products \(V_0\delta V\) and \(\delta V^2\) are defined on
\(\mathcal{D}(U)\) we may write
\[
q_{\lambda,t}(\psi, \phi) = (Q_{0,0}\psi,\phi) - \lambda^2(\psi,\phi) 
- t^2(\delta V^2\psi,\phi) + 2\lambda(V_0\psi,\phi) 
+ 2\lambda t(\delta V\psi,\phi) - 2t(V_0\delta V\psi,\phi). 
\]
Since this is valid for any \(\phi\) from \( \mathcal{D}(U)\) 
on which \(q_{\mu,t}\) is known to be closed the first representation 
theorem of Kato (\cite{Kt-66}) implies
\(\psi\in \mathcal{D}(Q_{\lambda,t})\) and 
\[
Q_{\lambda,t}\psi = Q_{0,0}\psi - \lambda^2\psi - t^2\delta V^2\psi
+ 2\lambda V_0\psi + 2\lambda t\delta V\psi - 2tV_0\delta V\psi.
\]
and by switching the roles of \(Q_{\lambda,t}\) and \(Q_{0,0}\) we have
\(\mathcal{D}(Q_{\lambda,t}) = \mathcal{D}(Q_{0,0})\) and the operator identity
\begin{equation}\label{Q_operator}
Q_{\lambda,t} = Q_{0,0} - \lambda^2 - t^2\delta V^2
+ 2\lambda V_0 + 2\lambda t\delta V - 2tV_0\delta V
\end{equation}
on \(\mathcal{D}(Q_{0,0})\).

In other words, as a function of \(t\), \(Q_{\lambda,t}\) is holomorphic
of type (A) as defined in \cite{Kt-66}.
Thus, the domain \(\mathcal{D} = \mathcal{D}_t 
= \mathcal{D}(Q_{\lambda,t})\)
is independent of both \(\lambda\) and \(t\).

We are now in a position to differentiate the quadratic eigenvalue equation
(\ref{QEP}) written in the form
\[
(\psi,Q_{\lambda,t}\phi) = 0
\]
where \(\psi = \psi(t) \in \mathcal{D}\), \(\lambda = \lambda(t)\), 
and \(\phi\) is any vector from  \(\mathcal{D}\) independent of \(t\)
and \(Q_{\lambda,t}\) is given by (\ref{Q_operator}) (note that for real \(\lambda\)
the operator \(Q_{\lambda,t}\) is selfadjoint).
By differentiating this using the Leibniz rule we obtain
\[
(\psi',Q_{\lambda,t}\phi) +(\psi,Q_{\lambda,t}\phi)') = 0
\]
and by (\ref{Q_operator})
\newpage
\[
(\psi',Q_{\lambda,t}\phi) =
\]
\[ - 2\lambda\lambda'(\psi,\phi) - 2t(\psi,\delta V^2\phi) 
+ 2\lambda'(\psi,V_0\phi) + 2\lambda' t(\psi,\delta V\phi) + 2\lambda (\psi,\delta V\phi)
- 2(\psi,V_0\delta V\phi) = 0.
\]
Now set \(\phi = \psi\). Using \(Q_{\lambda,t}\psi = 0\) we obtain
\[
\lambda'(\psi,(V_0 - \lambda + 2t\delta V)\psi) +
(\psi,(V_0 - \lambda + 2t\delta V)\delta V \psi) = 0.
\]
Hence
\begin{equation}\label{nu'1_linear}
\lambda' = \frac{(\psi,(V_0 - \lambda + t\delta V)\delta V \psi)}
{(\psi,(V_0 - \lambda + t\delta V)\psi)} =
\frac{(\psi,(V_t - \lambda I)\delta V \psi)}
{(\psi,(V_t - \lambda I)\psi)}
\end{equation}
which is in accordance with the heuristic formula (\ref{nu'1}).
Using that \(V_t - \lambda I\) and \(\delta V\) are positive semidefinite
and commuting we infer that \(\lambda'\) is non-negative. The positive
definiteness of the former  follows from \(\lambda < -M\) and our assumption
(\ref{order}).

\begin{figure}
\includegraphics[width=8cm]{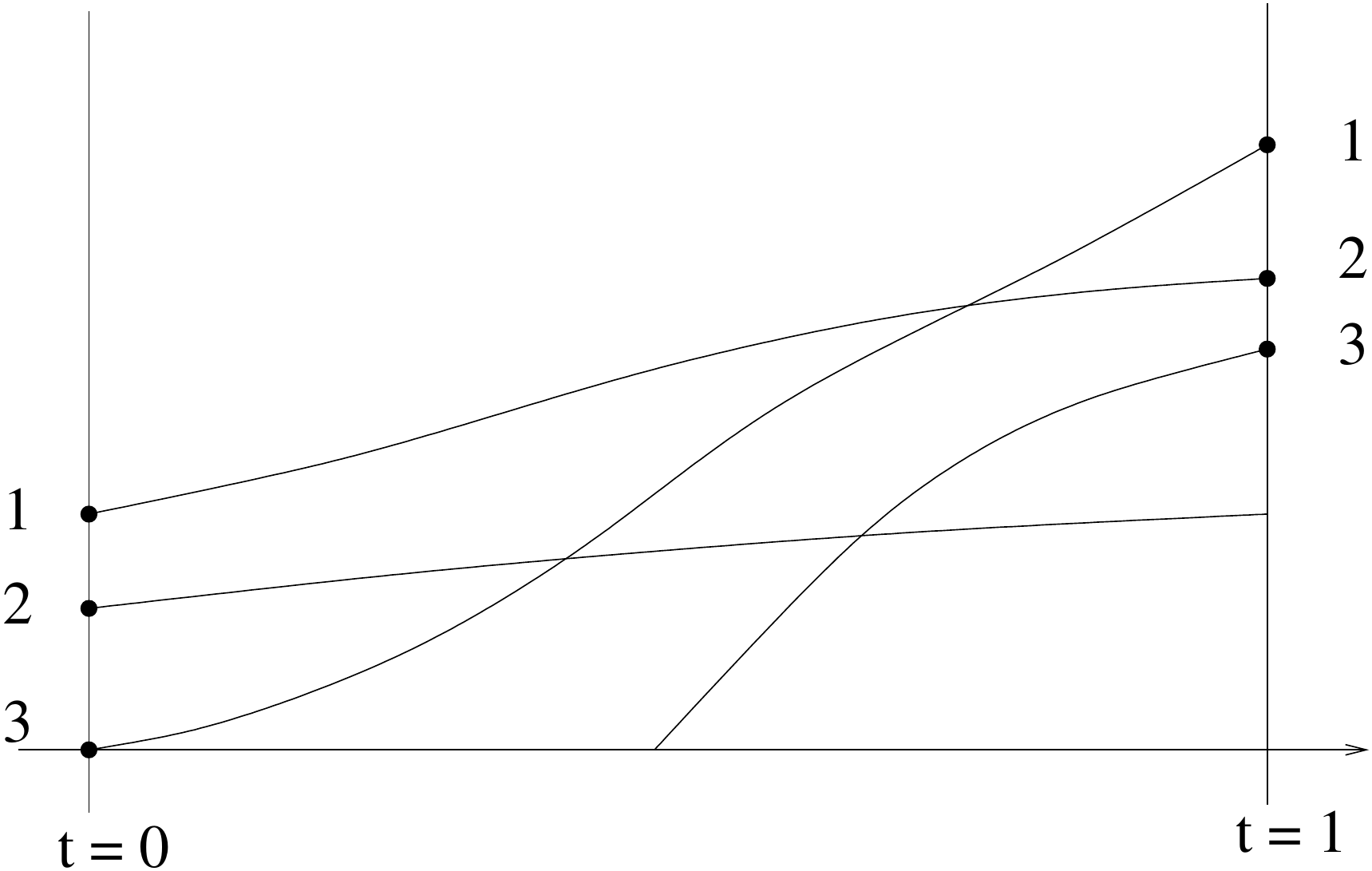}
\caption{Ordering ascending eigenvalues \label{crossing}}
 \label{fig1}
\end{figure}

We have thus obtained \(n\) functions
\[
\lambda_1(t) \geq \cdots  \geq \lambda_n(t)
\]
non decreasing in \(t\) and representing discrete eigenvalues
of \(K_t\) together with their multiplicities and such that
\[
\lambda_1(0) = \lambda_1^{(0)} \geq \cdots \lambda_n(0) = \lambda_n^{(0)}
\]
However this non-increasing ordering need not be kept by
\(\lambda_i(t)\) for all \(t\) because some other isolated 
(but also non-increasing in \(t\)) eigenvalues may be crossing 
so that from some
\(t\) onwards we may have more than \(n\) eigenvalues larger than
\(\tilde{\lambda}\). To overcome this inconvenience we sort the eigenvalues
\(\lambda_i(t)\) non-increasigly in \(i\) for any \(t\) thus obtaining
the \(n\) top eigenvalues
\[
\tilde{\lambda}_1(t) \geq \cdots \geq \tilde{\lambda}_n(t)
\]
 which are still continuous
and non-decreasing in \(t\) (but possibly only piecewise differentiable)
This situation is illustrated on Figure
\ref{crossing}. 

Now
\[
\lambda_i^{(1)} =\tilde{\lambda}_i(1)
\]
are the top eigenvalues of \(K_1\) as asserted in the statement.
Q.E.D.\\

Of course, a completely symmetric estimate holds for {\em plus} 
eigenvalues (just turn \(V\) into \(-V\)) which is the
common situation in Relativistic Quantum Mechanics.\\

\begin{figure}
\includegraphics[width=8cm]{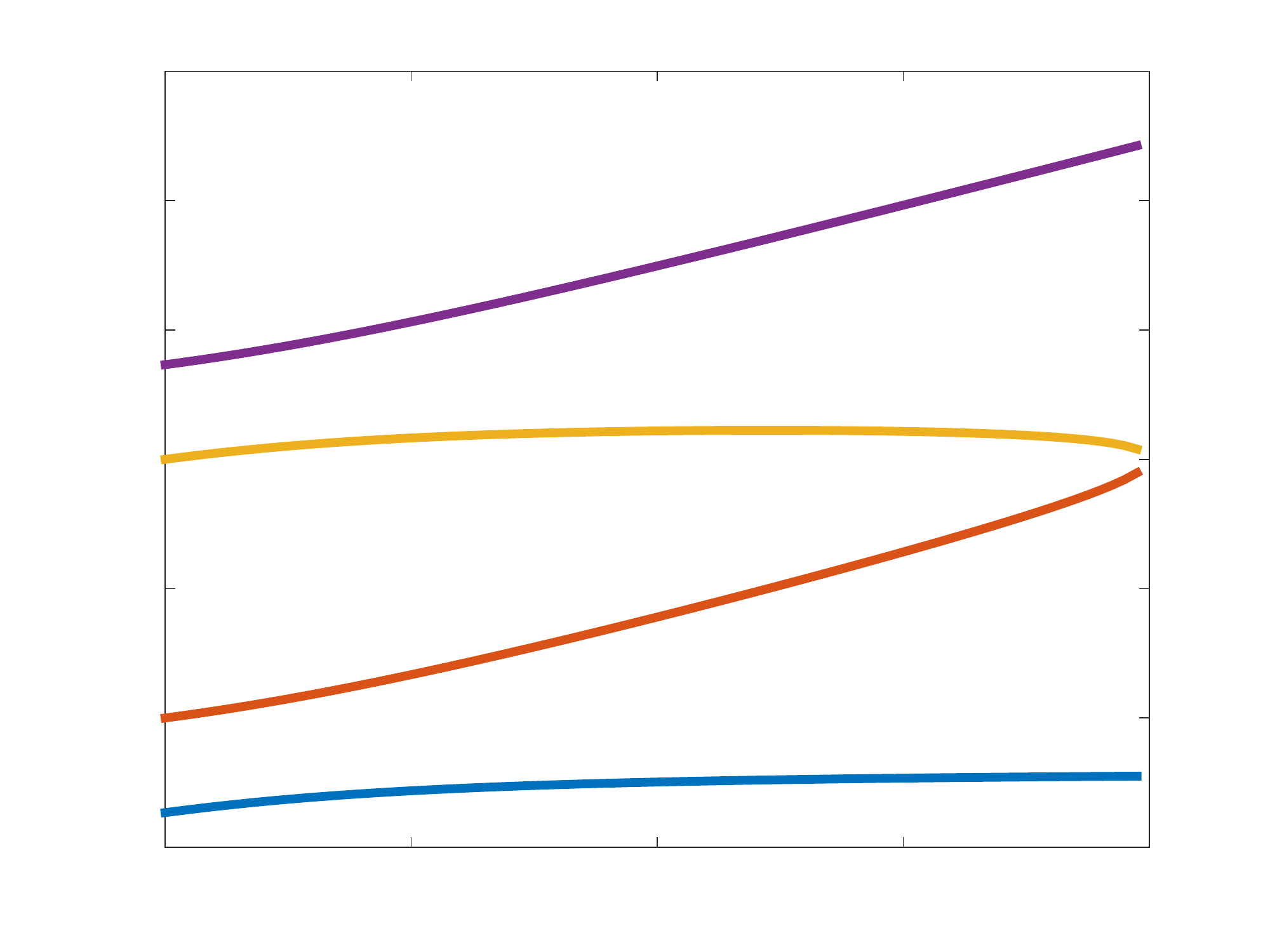}
\caption{Eigenvalues as functions of \(t \in [0,2]\) \label{nonmonotone}}
\end{figure}

The positive eigenvalues need not be monotone. 
The following example will show that. 
\begin{Example}
Set
\[
U^2 =
\bmat{cc}
2 & -1   \\
-1   &  2 \\
\emat,
\quad
V = t
\bmat{cc}
 1 & 0   \\
 0 & 0 \\
\emat
\]
with \(0 \leq t \leq 2\).
Note that here different \(tV\) 's commute. 

The non-monotone behaviour of \(\lambda_1^+\)
is shown on Figure \ref{nonmonotone}.
\end{Example}

In fact, as the figure indicates, at \( t = 2\)
both \(\lambda_1^-\) and \(\lambda_1^+\) are equal to
\(-1\) and for \(t > 2\) they will become non-real.
However, it should be noted that the non-monotonicity appears only
after a good while, as is also observed in \cite{SSW}.\\

There can well be cases, however, in which the unperturbed potential vanishes,
or is lumped together with the free Hamiltonian.
Then the mononicity will hold for both positive and negative part
of the spectrum in opposite directions, respectively --- as in the perturbation
of the Klein-Gordon oscillator below.   

\section{The eigenvalue bounds} Take the perturbation \(\delta V\) as bounded
and commuting with \(V\), then set \(\tilde{V} = V + \delta V\). Then
\[
(\psi,(V - \|\delta V\|I)\psi) \leq (\psi,\tilde{V}\psi) \leq 
(\psi,(V + \|\delta V\|I)\psi). 
\]
Now use the fact that the potential
and the eigenvalue enter (\ref{QEP}) only as a difference.
Since \(V \pm \|\delta V\|I\) produces the eigenvalues
\(\lambda_j \pm \|\delta V\|\)
the monotonicity  Theorem \ref{monotone1} is applicable and it implies
\begin{equation}\label{ll'}
\lambda_j - \|\delta V\|\leq \tilde{\lambda}_j \leq \lambda_j + \|\delta V\|.
\end{equation}
Here, of course, to insure monotonicity we must assume that \(\delta V\)
is sufficiently small as to have 
\begin{equation}\label{marg1}
V - \|\delta V\| I \geq -M.
\end{equation}
In the most interesting case of a positive potential this is insured,
if
\begin{equation}\label{marg2}
\|\delta V\|  \leq M
\end{equation}
which leaves ample a margin for any practical purpose.
More precisely, if
\[
\delta_\pm = {\sup_\psi \atop \inf_\psi}\frac{(\psi,\delta V\psi)}{(\psi,\psi)}
\]
then (\ref{ll'}) is strengthened into
\begin{equation}\label{ll'pm}
\lambda_j - \delta_-\leq \tilde{\lambda}_j \leq \lambda_j + \delta_+.
\end{equation}

This estimate can not be improved in general, just take \(V\) bounded
and \(\delta V\) a scalar multiple of \(V\).\\

\begin{center}
\begin{figure}
\includegraphics[width=14cm]{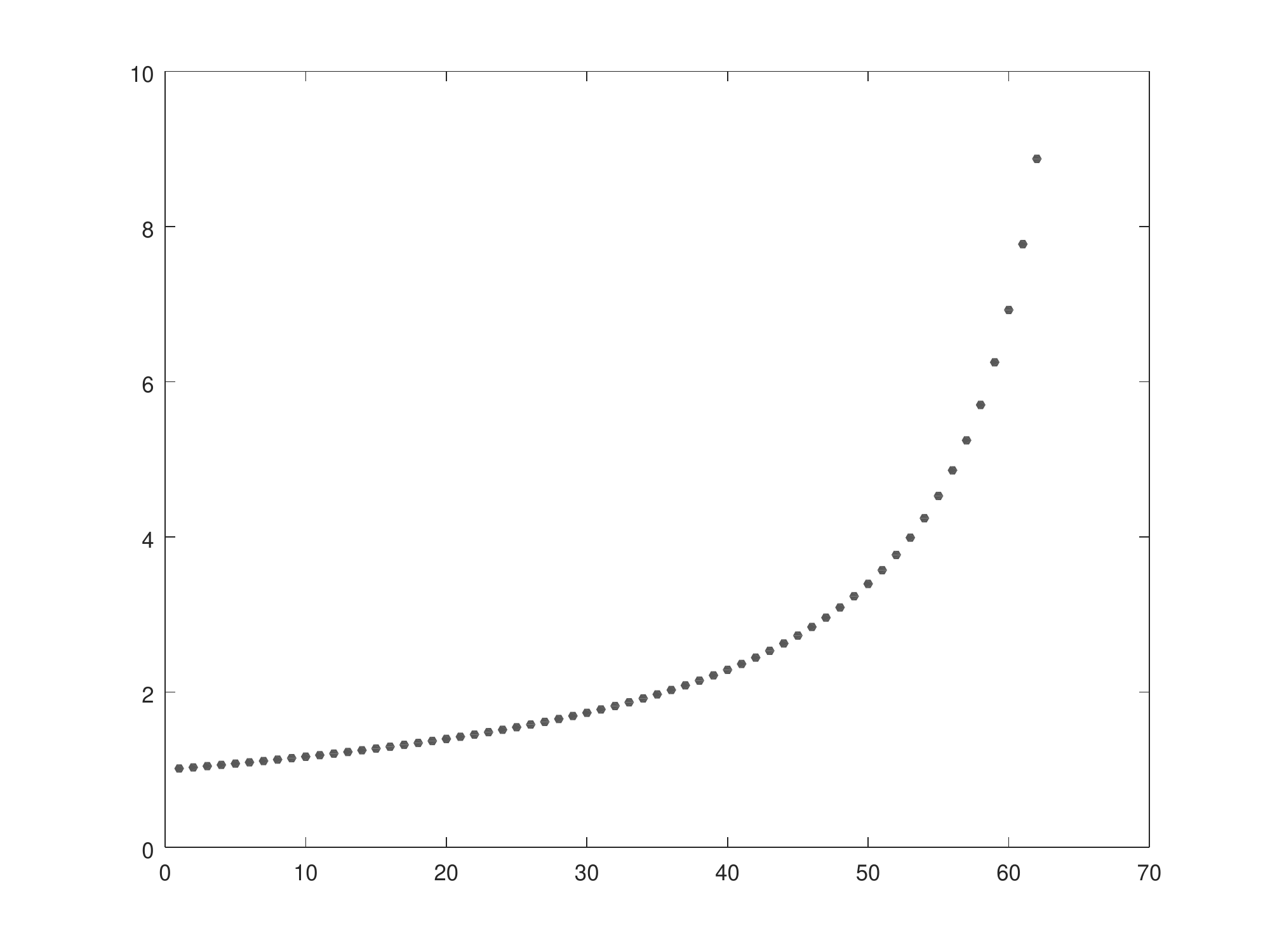}
\caption{Penalty as function of the atomic number \label{penalty}}
\end{figure}
\end{center}

We will compare our bound with the one obtained in 
\cite{NakicVeselic2020} which reads\footnote{For simplicity in 
the following we drop
the indices of the eigenvalues and also --- to prevent
moving of the essential spectrum --- we will assume all perturbations
\(\delta V\) to have compact support thus insuring relative compactness.}
\begin{equation}\label{bound_from_7}
|\tilde{\lambda} - \lambda| \leq \mbox{NVbound} := 
\frac{|\lambda|\|\delta VU^{-1}\|}{1 - \|VU^{-1}\|} \leq
\zeta\|\delta V\|.
\end{equation}
with the penalty
\begin{equation}\label{zeta}
\zeta = \frac{|\lambda|\|U^{-1}\|}{1 - \|VU^{-1}\|}.
\end{equation}
We will now compare this bound with the bound (\ref{ll'}) which is just 
\(\|\delta V\|\) on a concrete example. We 
evaluate the penalty \(\zeta\) for the case of the ground 
state of the {\em Coulomb relativistic Hamiltonian} given by
\begin{equation}\label{coulomb}
U^2 = -h^2c^2\left(\frac{\partial^2}{\partial x_1^2} +
\frac{\partial^2}{\partial x_2^2} +
\frac{\partial^2}{\partial x_3^2}\right)
+ m^2c^4,\quad
V = -\frac{Ze^2}{|x|}, 
\end{equation}
where \(m,h,c,e,Z\) are the common physical constants, in particular,
\(Z\) is the atomic number. The ground state energy is given as
\begin{equation}\label{ground}
\lambda_0 = mc^2\left[1 + \frac{Z^2\alpha^2}{(0.5 + 
\sqrt{.25 - Z^2\alpha^2)^2}}\right]^{-1/2}
\end{equation}
where \(\alpha = \frac{e^2}{hc} \approx 1/137\) is the fine structure constant.
Note that the ground state is always positive and so of plus type.
Obviously \(\|U^{-1}\| = 1/(mc^2)\) whereas by the known estimate
(see e.g.\ \cite{Kt-66} , Ch.\ V, (5.30)) we have
\[
\|VU^{-1}\| = 2Z\alpha
\]
and, by inserting in (\ref{zeta}) we obtain 
\[
\zeta = \left[1 + \frac{Z^2\alpha^2}{(0.5 + 
\sqrt{.25 - Z^2\alpha^2)^2}}\right]^{-1/2}/(1 - 2Z\alpha).
\]
The behaviour of this penalty as a function of the atomic number \(Z\)
is shown on Figure \ref{penalty}.
Thus our present bound outdoes the one from \cite{NakicVeselic2020}
significantly for larger atomic numbers.

There are, hovewer, examples where our theory is void and (\ref{bound_from_7})
still valid because the latter has no restriction to positive eigenvalues only.
Such is the case of deep potential well which may have negative 
plus-eigenvalues.

While the value \(\|\delta V U^{-1}\|\) is not easy to evaluate
with \(U^2\) Laplacian and \(\delta V\) a multiplication operator, we can do
with another 
relative estimate (we keep with non-positive \(V\))
\begin{equation}\label{Vrel}
-\gamma_-(V\psi,\psi) \leq (\delta V\psi,\psi) \leq -\gamma_+(V\psi,\psi)
\end{equation}
which is easily verifiable, being a consequence of
\begin{equation}\label{Vrelx}
-\gamma_-V(x) \leq \delta V(x) \leq -\gamma_+V(x).
\end{equation}

This gives
\[
(1 - \gamma_-)(V\psi,\psi) \leq ((V + \delta V)\psi,\psi) 
\leq (1 - \gamma_+)(V\psi,\psi).
\]
This leads to an interesting bound which uses not only the unperturbed eigenvalue 
\(\lambda\) and the perturbation \(\delta V\) but the eigenvalues \(\lambda(\varepsilon)\)
belonging to the potential \(\varepsilon V\) as follows
\begin{equation}\label{ll'rel}
\lambda(1 - \gamma_+) \leq \tilde{\lambda} \leq \lambda(1 - \gamma_-)
\end{equation}
where \(\lambda = \lambda(1)\). So, the perturbed eigenvalue \(\tilde{\lambda}\)
is contained in the interval
\begin{equation}\label{interval}
\mathcal{I} = [\lambda(1 - \gamma_-),\lambda(1 - \gamma_+)]
\end{equation}
(note that \(\lambda(\varepsilon)\) is falling with \(\varepsilon\)).

The obtained bound is also sharp as is seen by taking \(\delta V\)
proportional to \(V\) in which case \( \gamma_\pm\) are just equal.\\

The knowledge of the whole family \(\lambda(\varepsilon)\) is seldom 
available explicitly
but just in the Coulomb case we have the explicit formula (\ref{ground}). Thus,
if the perturbing potential satisfies (\ref{Vrel}) with \(V\) from
(\ref{coulomb})
then for the ground state the perturbed eigenvalue 
\(\tilde{\lambda}\) satisfies 
(\ref{ll'rel}) where by (\ref{ground})
\[
\lambda(\varepsilon) = mc^2\left[1 + \frac{\varepsilon^2 Z^2\alpha^2}{(0.5 + 
\sqrt{.25 - \varepsilon^2 Z^2\alpha^2)^2}}\right]^{-1/2}.
\]

To illustrate the power of this  kind of estimate we take as the perturbation

\begin{equation}\label{dV_coulomb}
\delta V(x) = \tau\left\{
\begin{array}{rr}
\frac{Ze^2}{|x|},& |x| > l\\
{} \\
\frac{Ze^2}{l},  & |x| \geq  l,
\end{array}
\right.\quad 0 < \tau < 1.
\end{equation}
This perturbation has certain physical appeal, so we will go in some detail.
We have here
\[
\delta_- = 0,\quad \delta_+ = \frac{Ze^2}{l}
\]
with
\[
\lambda \leq  \tilde{\lambda} \leq \frac{Ze^2}{l}
\]
and
\[
\gamma_- = 0,\quad \gamma_+ = \tau
\]
with
\[
\lambda \leq \tilde{\lambda} \leq \lambda(Z(1 - \tau)).
\]

\begin{center}
\begin{figure}
\includegraphics[width=17cm]{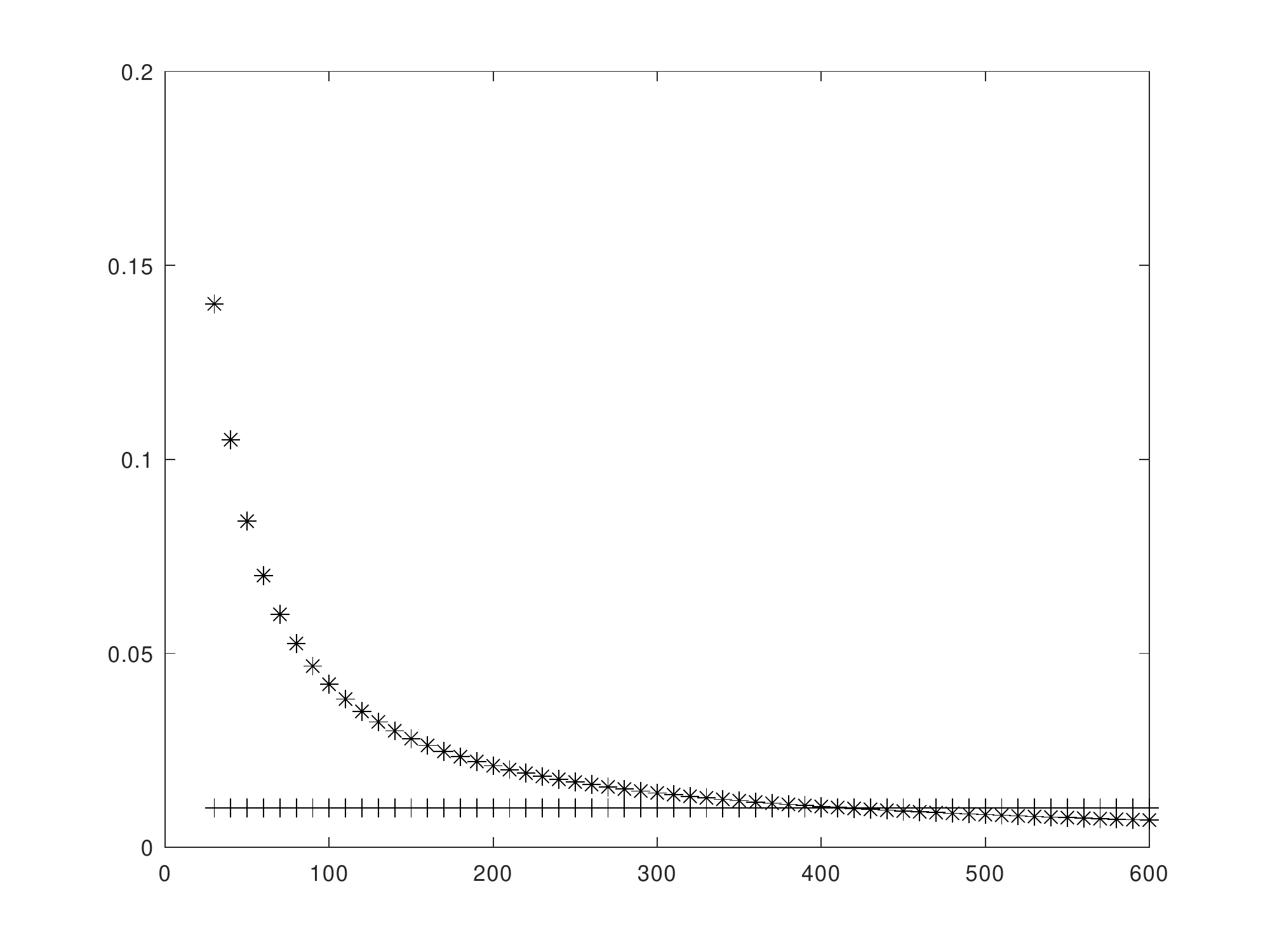}
\caption{Absolute (*) and relative bound (+)\label{2bounds}}
\end{figure}
\end{center}

For \(Z = 40\) the behaviour of both bounds as functions
of \(l\) is shown on Figure \ref{2bounds}
The new bound (represented by crosses) is independent of the radius \(l\)
and is clearly better as the norm bound (starred line) except for very
large radii (from some 400 classical electron radii onwards). Besides, the value
of the new bound is some \(0.01\). That is, \(1 \%\) change of the potential
produces \(0.1 \%\) change of the eigenvalue. Thus the relative
perturbation (\ref{Vrelx}) produces very small changes of the eigenvalue.
There is no reason to believe that this property holds only for the few
potentials for which the dependence \(\lambda(\varepsilon)\) 
is explicitly known.

Of course the two bounds (\ref{ll'pm}) and (\ref{ll'rel}) can be combined
into one: if
\[
\delta_- - \gamma_-V(x) -\leq  \delta V(x) \leq \delta_+ - \gamma_+V(x)
\]
then
\begin{equation}\label{ll'comb}
\delta_- +\lambda(1 - \gamma_+) \leq \tilde{\lambda} \leq \delta_+ + 
\lambda(1 - \gamma_-).
\end{equation}

\begin{Remarks}
\begin{enumerate}
\item Note that our key results, notably Theorem \ref{monotone1} and the bounds
(\ref{ll'}), (\ref{ll'pm}, (\ref{ll'rel}) give not merely relations between
existing eigenvalues, they rather include the {\em existence statements}, so
for instance, in (\ref{ll'rel})
between each pair \(\lambda(1 - \gamma_-), \lambda(1 - \gamma_-)\) there is
a perturbed \(\tilde{\lambda}\) (including multiplicities).
\item Note the remarkable fact: under {\em just any} perturbation \(\delta V\)
satisfying (\ref{Vrelx})
no eigenvalue can cross the boundary \(mc^2\) of the continuous spectrum,
what is more, {\em the perturbed eigenvalues will be infinitely many and also accumulate at
this boundary.} This is a general property of locally deformed
Coulomb potentials.

\item The estimates of the type (\ref{ll'rel}) will obviously hold for other
quantum mechanical Hamiltonians (Schroedinger, Dirac...) where the eigenvalues are
known to depend monotonically on the potential.
\end{enumerate}
\end{Remarks}

Our next example will the Klein-Gordon oscillator.
Here we have
\begin{equation}\label{U_osc}
U = m^2c^4 + m^2c^2\omega^2|x|^2 ),\quad V(x) = \alpha x_1
\end{equation}
with
\[
x = 
\begin{bmatrix}
x_1  \\
\vdots\\
x_n \\
\end{bmatrix}
\in\mathbb{R}^n,\quad p = -ih\nabla.
\]
Here \(\omega > 0\) is the classical oscillator frequency.
The model is explicitly solvable and is readily seen that in this case
\[
a = \|VU^{-1}\| = \frac{|\alpha|}{mc\omega},
\]
whereas the spectrum is discrete and given by the eigenvalues
\begin{equation}\label{mu_harm}
\lambda_k^\pm = \pm\sqrt{(2mc^2h\omega\nu_k(a) + m^2c^4)(1 - a^2)},
\quad k = [k_0,\ldots,k_n],\quad k_i = 0,1,2,\ldots
\end{equation}
with 
\[
\nu_k(a)= (1 - a^2)^{1/2}\left(k_1 + \frac{1}{2}\right) + k_2 + \frac{1}{2}
+ \cdots + k_n + \frac{1}{2},\quad k_1,\ldots,k_n \geq 0
\]
which are simple for \(n = 1\). We see that the case \(a \neq 0\) is
completely out of reach of our theory because the homogeneous field
\(V = \alpha x_1\) is deeply indefinite. In contrast, the case
\(a = 0\) --- this is the pure Klein-Gordon oscillator --- we have
non-trivial in fact, quite strong results. Then in the key formula 
(\ref{nu'1_linear}) we have \(V_0 = 0\) and it reads
\begin{equation}\label{nu'0_linear}
\lambda' = \frac{(\psi,(- \lambda + t\delta V)\delta V \psi)}
{(\psi,(- \lambda + t\delta V)\psi)}. 
\end{equation}
So, the condition (\ref{order}) is automatically fulfilled
if \(V_1 = \delta V\) is positive semidefinite. Consequently,
{\em plus and minus eigenvalues are monotone (in opposite directions)
and our eigenvalue bounds (\ref{ll'}) and (\ref{ll'pm})  
hold without any restriction}.

We can again compare this with the bound (\ref{bound_from_7}) where
\(V = 0\) gives
\[
|\tilde{\lambda} - \lambda| \leq \frac{|\lambda|\|\delta V\|}{M}
\]
which is as good as ours for the ground state \(\lambda = \lambda_0^-
= -M\). For higher energies
the new bound gets better and better.

The estimate (\ref{ll'rel}) can be used here, too. Perturbations,
satisfying (\ref{Vrelx})  will yield potentials \(V + \delta V\)
whose eigenvalues will have {\em the same spectral asymptotics as
the Klein-Gordon oscillator}. Combined bounds (\ref{ll'comb})
will hold, as well.\\

In trying to relax the conditions of Theorem \ref{monotone1}
we will use the minimax formula obtained in \cite{LangerTretter2006}. 
We set
\begin{equation}\label{p+-}
p_\pm(\psi) = 
(\psi,V\psi) \pm \sqrt{(\psi,V\psi)^2 + (U\psi,U\psi) - 
(V\psi,V\psi)},\quad \|\psi\| = 1,\quad
\psi \mathcal{D}(U).
\end{equation}
Then, as is readily seen
\[
\min\sigma_+(K) = \inf p_+(\psi),\quad \max\sigma_-(K) = \sup p_+(\psi)
\]
and, as proved in \cite{LangerTretter2006}
\begin{equation}\label{minimax}
\lambda_k^+ = \min_{S_k}\max_{\psi\in S_k,\|\psi\|=1}p_+(\psi),\quad
\lambda_k^- = \max_{S_k}\min_{\psi\in S_k,\|\psi\|=1}p_-(\psi).
\end{equation}
where \(S_k\) is any \(k\)-dimensional
subspace of \(\mathcal{D}(U)\) and \(\lambda_i^\pm\) are the inner discrete eigenvalues with multiplicities 
ordered as
\[
\cdots \leq \lambda_2^- \leq \lambda_1^- < 
\lambda_1^+ \leq \lambda_2^+ \leq \cdots \qquad \mathcal{I} = 
(\lambda_1^-, \lambda_1^+).
\]
This time our conditions should accomodate the fact that, while working with
positive potentials \(V_0,V_1\) we will be considering 
{\em minus eigenvalues} which
can well become positive if the potentials are deep enough (e.g. 
deep potential wells). So, the assumption \(\|V_0\psi\| \leq \|V_1\psi\| < 1\)
would not do and we have to use (\ref{condition_on_V}). 
In order to prove the monotonicity we will have to investigate
the dependence of \(p_\pm\) on \(V\), so we shall write
\[
p_\pm(\psi) = p_\pm(\psi,V).
\]
If we prove the monotonicity of \(V \mapsto p_+(\psi,V)\) for any
fixed unit \(\psi\) then the minimax formula (\ref{minimax})
immediately implies the same for the plus eigenvalues.
By setting
\[
v = (\psi,V\psi),\quad w = (V\psi,V\psi), \quad \chi = (U\psi,U\psi)
\]
we have to study the behaviour of the function
\[
v,w \mapsto f(v,w) = p_-(\psi,V)= v - \sqrt{v^2 + \chi - w}
\]
on the domain \(v \geq 0,\quad 0 \leq w < \chi\).
We have
\[
\frac{\partial f}{\partial v} =
1 - \frac{v}{\sqrt{v^2 + \chi - w}} > 0,
\]
\[
\frac{\partial f}{\partial w} =
 \frac{1}{2(\sqrt{v^2 + \chi - w})} > 0.
\]
So, if \(V^2\) is growing, then both \(v\) and  \(w\) will be growing,
 hence \(f\) is growing for any fixed 
\(\psi\) that is, if \(V_0, V_1 \geq 0\) and \(V_0^2 \leq V_1^2\) then
\[
p_-(\psi,V_0) \leq p_-(\psi,V_1).
\]
Thereby this growing goes over to the minimax quantities from (\ref{minimax}).

Now go into the proof of Theorem \ref{monotone1}. There for any
\(t \in [0,1]\) we have the top eigenvalues, counting multiplicities
\[
\tilde{\lambda}_1(t) \geq \cdots \geq \tilde{\lambda}_n(t)
\]
of \(\sigma_-(K)\), piecewise differentiable in \(t\).
By the minimax formula we infer to the waxing in \(t\) as well.
Altogether we have
\begin{teo}\label{monotone2} 
Let \(V_0,V_1\) be symmetric and and positive semidefinite 
with \((V_1\psi,\psi) \leq (V_1\psi,\psi),\quad \psi \in \mathcal{D}(U)\)
and and let
\begin{equation}\label{condV_mu}
\|(V_0 -  \mu)\psi\|  < 1,\quad
\|(V_1 -  \mu)\psi\| < 1, \quad \psi \in \mathcal{D}(U)
\end{equation}
for some real \(\mu\).
Then the assertions of Theorem \ref{monotone1} hold true.
\end{teo}

It is instructive to compare the conditions and also the results
of this theorem with those of Theorem \ref{monotone1}. The commutativity
is dropped, and the clumsy assumptions on products of \(V_1,V_2\) 
is replaced by
(\ref{condV_mu}). If \(V_1\) (and then also \(V_0\)) is bounded, then
(\ref{condV_mu}) is implied by the single inequality
\begin{equation}\label{condV_norm}
\|V_1\| < 2\|U^{-1}\|^{-1}.
\end{equation}
However interesting in its own sake this theorem is by itself
not sufficient to produce two-sided eigenvalue bounds
(\ref{ll'}), (\ref{ll'pm}) and (\ref{ll'rel}). (The last one will
still hold 
under additional assumption that
{\em both} \(V\) and \(V + \delta V\) are non-negative.) 
This calls for some further research.

On the other hand Theorem \ref{monotone1} appears to be a special case
of Theorem \ref{monotone2}. But a closer look at the proof of the former
shows that it in fact holds in {\em any} gap of the essential spectrum
and not only in that around \(\mathcal{I}\) which Theorem 
\ref{monotone1} does not.
Also one might ask why in proving Theorem \ref{monotone2} we did not use 
minimax formulae alone without recurring to analytic perturbation theory.
The reason is that as far as yet it is only the latter that guarantees
the existence of the perturbed eigenvalues -- at least with the present
state of minimax theory as presented in \cite{LangerTretter2006}.

A much stronger, in fact, the maximal result follows for the important special case of
\[
V_t = tV.
\]
We have
\begin{cor}\label{proportional_t}
If \(V\) is positive semidefinite then with the potential \(tV, \ t > 0\) 
the minus eigenvalues are monotone 
(as functions
of \(t\)) in the sense of Theorem \ref{monotone2}. This state of affairs
carries on as long as \(\|(tV - \mu I)U^{-1}\| < 1\) for some \(\mu\).
\end{cor}
{\bf Proof.} We have
\[
p_-(\psi) = p_-(\psi,t) =
t(\psi,V\psi) - \sqrt{t^2(\psi,V\psi)^2 + (U\psi,U\psi) - 
t^2(V\psi,V\psi)}
\]
and hence
\[
p_(\psi,t)' = (\psi,V\psi) - t\frac{(\psi,V\psi)^2 - (V\psi,V\psi)}
{\sqrt{t^2(\psi,V\psi)^2 - t^2(V\psi,V\psi) + (U\psi,U\psi)}}
\]
\[
= 
(\psi,V\psi) + t\frac{(\|(V - (\psi,V\psi)I)\psi\|^2}
{\sqrt{t^2(\psi,V\psi)^2 - t^2(V\psi,V\psi) + (U\psi,U\psi) }}
\geq 0
\]
which holds as long as the radicand above is positive that is,
if \(\|(tV - \mu I)U^{-1}\| < 1\) for some \(\mu\). 

The rest of the
proof is as in Theorems \ref{monotone1}, \ref{monotone2} respectively.
Q.E.D.\\

{\bf Finite matrices}. All our results above naturally comprise 
the case of finite
matrices \(U,V\). This is, however, of limited use in a computational
environment where by various approximations and errors the 
commutativity of different \(V\)-s is easily lost. And 
without commutativity there
is no way to convert the Loewner's theorem -- monotonicity of positive
operators does not generally extend to their squares -- and this is needed to
infer to the monotonicity of \(\mu\) in (\ref{nu'1}).\\

{\small {\bf Acknowledgement.} The author would like to thank I. Naki\'{c},
Zagreb, for helpful discussions.}


\begin{thebibliography}{ABC-99x}
\bibitem{CurgusNajman95} B. \'Curgus, B. Najman,
Quasi-uniformly positive operators in Krein space,
in {\em Operator Theory: Advances and Applications, Vol. 80} (1995) 90-99.
\bibitem{glr} Gohberg, I., Lancaster, P., Rodman, L., {\em Matrices and
   indefinite scalar products}, Birkh\"{a}user, Basel 1983.

\bibitem{Hall1} Hall, RL. Relativistic comparison theorems, 
Phys. Rev. A (3) 81 (2010), no. 5, 052101, 4 pp.

\bibitem{Hall2} Hall, RL. and Zorin, P. Sharp comparison theorems for the 
Klein–Gordon equation in d dimensions, arxiv:1509.01728v2 (2016).
\bibitem{Hall3} R. L. Hall and M. D. Aliyu, Comparison theorems for 
the Klein-Gordon equation in  dimensions, Phys. Rev. A 78, 052115 (2008).
\bibitem{Kt-66} Kato, T., {\em Perturbation theory for linear operators},
   Springer, Berlin 1966.
\bibitem{LangerTretter2008}
        Langer, Heinz and Najman, Branko and Tretter, Christiane, Spectral theory of the
Klein-Gordon equation in Krein spaces
        Proc. Edinb. Math. Soc. {\bf 51} (2008) 711-750.
\bibitem{Langer82}Langer, Heinz, Spectral functions of definitizable 
operators in Krein spaces, Springer Lecture Notes in 
Mathematics {\bf 948} (1982).

\bibitem{LangerTretter2008-P}
Langer, Heinz and Najman, Branko and Tretter, Christiane, 
Spectral theory of the 
Klein-Gordon equation in Pontrjagin spaces,
        Commun. Math. Phys. 267, 159-180 (2006)

\bibitem{LangerTretter2006} Langer, Matthias and Tretter, Christiane,
Variational principles for eigenvalues of the Klein--Gordon equation, 
Journal of mathematical physics {\bf 10} (2006) 103506.

\bibitem{Najman1980} Najman, B., Spectral properties of the operators of 
Klein-Gordon type, Glasnik Mat. Ser. III 15(35) (1980), no. 1, 97–112.
\bibitem{Najman1983} Najman, B.,
Eigenvalues of the
Klein-Gordon equation
        Proc. Edinb. Math. Soc. {\bf 26} (1983) 181-190.

\bibitem{SSW} Schiff. L-I., Snyder, H., Weinberg, J., 
On the existence of stationary
states in the mesotron fields, Phys. Rev. {\bf 57} (1940) 315-318.
\bibitem{NakicVeselic2020} Naki\'{c}, Ivica; Veseli\'{c}, Kre\v{s}imir;
Perturbation of eigenvalues of the Klein-Gordon operators,
Rev. Mat. Complut. 33 (2020), no. 2, 557-581.

\bibitem{v0old} Veseli\'{c}, K., A spectral theory for the Klein-Gordon
equation with an electrostatic potential, Nucl. Phys.
{\bf A147} (1970) 215-224.

\bibitem{Vespseudo} K. Veseli\'{c}, Perturbation of pseudoresolvents 
and analyticity in 1/c in relativistic Quantum Mechanics, Communs. Math. Phys. 22, (1971) 27-43.

\bibitem{v0} Veseli\'{c}, K., On spectral properties of a class of
$J$-selfadjoint operators I, {\em Glasnik Mat.}, 7(27):229-248 (1972).

\bibitem{v0new} Veseli\'{c}, K., A spectral theory of the Klein-Gordon
equation involving a homogeneous electric field J. Operator Th.
{\bf 25} (1991) 319-330.

\bibitem{Veselic1983} Veseli\'{c}, K.,
On the nonrelativistic limit of the bound states of the Klein-Gordon equation.
J. Math. Anal. Appl. 96 (1983), no. 1, 63–84. 


\bibitem{Veselic-selfa} K. Veseli\'{c}, Spectral perturbation bounds for selfadjoint operators, Operators and Matrices {\bf 2} (2008) 307-339.


\end{thebibliography}
\end{document}